\documentclass[twocolumn]{aastex61}
\usepackage{epsfig}
\newcommand{\gsim}{\hbox{\rlap{$^>$}$_\sim$}}

\begin{document}
\title{Origin Of The Far Off-Axis GRB171205A}

\author{Shlomo Dado}
\affiliation{Physics Department, Technion, Haifa 32000, Israel}
\author{Arnon Dar}
\affiliation{Physics Department, Technion, Haifa 32000, Israel}

\begin{abstract} 
We show that observed properties of the low luminosity GRB171205A 
and its afterglow, like those of most other low-luminosity (LL) gamma ray 
bursts (GRBs) associate with a supernova (SN), indicate that it is an ordinary 
SN-GRB, which was produced by inverse Compton scattering of glory light by 
a highly relativistic narrowly collimated jet ejected in a supernova
explosion and viewed from a far off-axis angle. As such,  VLA/VLBI 
follow-up radio observations of a superluminal displacement of its 
bright radio afterglow from its parent supernova, will
be able to test clearly whether it is an ordinary SN-GRB viewed from 
far off-axis or it belongs to a distinct class of GRBs, which are
different from ordinary GRBs, and cannot be explained by
standard fireball models of GRBs as ordinary GRBs 
viewed from far off-axis.
\end{abstract}

\section{Introduction} 
The standard fireball model of gamma ray bursts (GRBs) cannot explain a 
difference of 6 orders of magnitude in luminosity between ordinary GRBs 
such as 130427A and nearby low luminosity (LL) GRBs, such as 980425 and 
060218, if they belong to the same population of GRBs, as indicated by 
their association with similar supernovae (e.g., Melandri et al.2014). 
Neither could it explain the much larger GRB production rate relative to 
the star formation rate of LL GRBs than that of ordinary 
GRBs. This led many authors to propose that ordinary GRBs and  
LL GRBs belong to different populations of GRBs (e.g., Toma 
et al. 2007; Liang et al. 2007; Guetta \& Della Valle 2007; Virgili et al. 
2009; Bromberg et al. 2012), which has been widely accepted.

In contrast, in the cannonball model of GRBs,  LL  SN-GRBs 
and LL SN-less GRBs are ordinary GRBs observed from far off-axis viewing 
angles (Dar \& De R\'ujula, 2000a, 2004; Dado, Dar \& De R\'ujula 
2003, 2009; Dado \& Dar 2017) whose rates depend on viewing angle and  detection 
threshold. When their effects are properly included, the extracted GRB 
rate from the observed rate as function of redshift becomes 
proportional to the star formation rate at all redshifts (Dado \& Dar 2014).
This conclusion, however, for very small redshifts was based on very few 
LL SN-GRBs and LL SN-less GRBs. 

Recently the LL GRB171205A, was detected 
by the Swift Burst Alert Telescope (BAT) (D'Elia et 
al. 2017). It was located by the Swift X-ray Telescope (XRT) in a 
nearby bright spiral galaxy (Izzo et al. 2017a) of redshift $z=0.0368$ 
(Izzo et al. 2017b). Assuming a standard cosmology model with $H_0 = 70$ 
km/s/Mpc, $\Omega_M = 0.30$, and $\Omega_\Lambda = 0.70$, the burst 
isotropic energy release Eiso as measured by Konus Wind (Frederiks et 
al. 2017) was $2.4\times 10^{49}$ erg in the 20-1500 keV range. So far, only 
limited evidence for the association of GRB171205A with a 
bright SN was reported (Ugarte Postigo et al. 2017; Cobb 2017). 
However, the observed properties of GRB171205A and its 
afterglow can be used already to further test whether it was an 
ordinary SN-GRB viewed from a far off-axis angle, 
or was it a member of a different population of GRBs.

\section{LL GRBs in the CB model}
In the cannonball (CB)  model (Dar \& De R\'ujula 2004) of 
GRBs, ordinary long duration GRBs are produced 
by inverse Compton scattering (ICS) of glory light (a light halo 
surrounding the GRB site) by highly relativistic jets of plasmoids 
(CBs) of ordinary matter (Shaviv \& Dar 1995) launched in stripped 
envelope supernova explosions (Dar et al. 1992) and in phase 
transition of neutron stars to quark stars (Shaviv \& Dar 1995; 
Dado \& Dar 2017). In the CB model, ordinary GRBs are viewed from 
an angle $\theta\approx 1/\gamma$ relative to the jet direction of 
motion, while low luminosity (LL) GRBs and X-ray flashes (XRFs) are 
ordinary GRBs viewed from far off-axis, $\gamma^2 \theta^2 \gsim 9$. 
Both ordinary GRBs and LL GRBs were predicted to display simple 
kinematical correlations between their main properties (Dar \& De 
R\'ujula 2000a). E.g., the peak energy of their time integrated 
energy spectrum satisfies $(1+z)\,Ep\propto \gamma\,\delta$, while 
their isotropic equivalent total gamma ray energy satisfies 
$Eiso\propto \gamma \delta^3$, where $z$ is their redshift and 
$\delta=1/\gamma(1-\beta\, cos\theta)$ is their Doppler factor. 
Hence, ordinary GRBs that are mostly viewed from an angle 
$\theta\approx 1/\gamma$, were predicted  
to satisfy 
\begin{equation}
(1+z)Ep\propto [Eiso]^{1/2},
\end{equation}
while far off-axis ($\theta^2\gamma^2>>1$) GRBs and SHBs were
predicted to satisfy (Dar \& De R\'ujula 2000a, Eq.(40))
\begin{equation}
(1+z)Ep\propto [Eiso]^{1/3}.
\end{equation} 
Eqs.(1) and (2) are compared in Figure 1 and 2 to the observational data 
on GRBs with known $z$, $Ep$ and $Eiso$. As can be seen from Figure 1, the 
CB model predicted correlation for ordinary GRBs as given by Eq.(1), which 
was later discovered empirically by Amati et al. (2002) and repeatedly 
tested by Amati et al. (2006, 2008, 2009, 2013) and by many other authors, 
is well satisfied by ordinary ($\theta\sim 1/\gamma$) GRBs. As can be seen 
from Figure 2, Eq.(2), the CB mode predicted correlation for far 
off-axis GRBs ($\theta\gamma\gg 1$), is also well satisfied by the LL-GRBs.
\begin{figure}[]
\centering
\epsfig{file=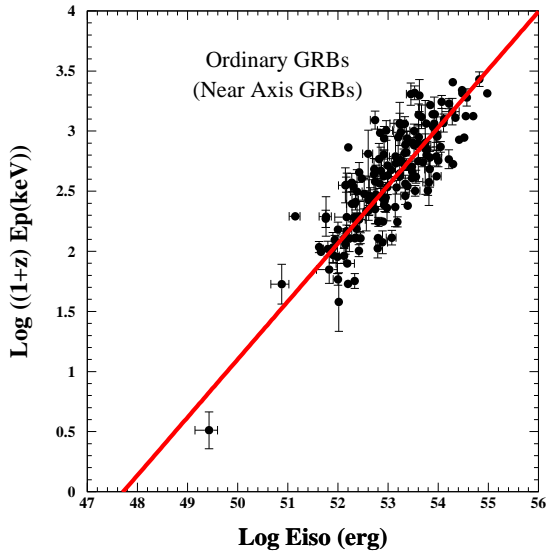,width=8.cm,height=8.cm}
\caption{The peak energy in the GRB rest frame as a function of
the isotropic equivalent total gamma ray
energy of ordinary GRBs viewed near axis. The line is the
correlation predicted by the CB model as given by Eq.(1).}
\label{Fig1}
\end{figure}  
\begin{figure}[]
\centering
\epsfig{file=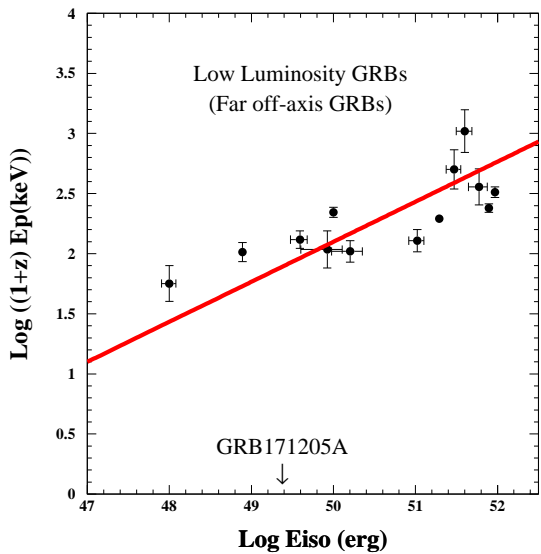,width=8.cm,height=8.cm}
\caption{The peak energy in the GRB rest frame as a function of
the isotropic equivalent total gamma ray
energy $Eiso$ of low luminosity (far off-axis) GRBs. The
line is the correlation predicted by the CB model, as given by
Eq.(2)) for far off-axis (low luminosity) GRBs.}
\label{Fig2}
\end{figure}
Eiso of the LL GRB171205A is indicated in Figure 2 
by an arrow. From the  correlation shown
in Figure 2 we estimate its $Ep\approx 79$ keV, and from the CB 
model relation (e.g., Dar \& De R\'ujula  2004) 
\begin{equation}  
Ep={\gamma\,\delta\, \epsilon \over (1+z)}\approx 
   {2\,\epsilon \over \theta^2 (1+z)}
\label{Eq3}
\end{equation}
for $\gamma^2\theta^2\gg 1$ and a mean energy of glory photons, 
$\epsilon \sim 1$ eV, we estimate a far off-axis viewing angle 
$\theta \approx 5\times 10^{-3}$, which we 
shall assume hereafter in our analysis of the observations of
GRB171205A.  

\section{Pulse Shape}
In the cannonball (CB) model (Dar \& De R\'ujula 2004), the production 
mechanism of GRBs is ICS of photons of a glory (light fireball) of radius 
$R$ surrounding the launching site of a highly relativistic jet of 
plasmoids (bulk motion Lorentz factor $\gamma\sim 10^3)$. The glory has a 
bremsstrahlung spectrum of temperature $T$, which yields a 
FRED (fast rise exponential decay) pulse shape of 
photons above a detection threshold $E_{min}$,
\begin{equation}
N(t,E>E_{min})\propto {t^2\over (\Delta^2+t^2)^2}\exp(-t/\tau)
\label{Eq4}
\end{equation}
where $\tau(E_{min})=(R/\gamma\,c)(kT\,E_{min})$ and $\Delta$ are best fit 
parameters. 
Eq.(4) is a simple interpolation between the early-time $\propto t^2$
rise due to the increasing effective cross section of the CB 
$\pi\, R_{CB}^2\propto t^2$ while it is crossing inside the 
glory region where the ambient photons are isotropic and 
the late-time behavior $\sim (1/t^2)\,exp(-t/\tau)$ 
when the CB is at a distance
$r=\gamma\,\delta\, c\, t/(1+z)\gg R$ from its launch site, 
whose derivation is as follows: 

At distances $r\gg R$ the number density of ambient photons intercepted 
by the CB decreases with distance as $1/r^2\propto t^{-2}$ and hence 
\begin{equation} 
{dn\over d\epsilon}\propto 
t^{-2}\,exp(-\epsilon/kT)/\epsilon. 
\label{Eq5} 
\end{equation}
In the CB's rest frame, the longitudinal momentum of the intercepted 
photons is reduced by a factor $1/2\gamma$.
Thus, in that frame, the photon's parallel momenta are
negligible compared to their transverse momenta, un-changed by the Lorentz 
boost. Let $b$ be the transverse 
distance of an emitted photon relative to the CB's direction 
of motion, which is intercepted at $r\gg R$. 
Its energy in the CB rest frame becomes $\epsilon'\approx \epsilon\,b/r$  
and after ICS in the CB it arrives with 
$E\approx \epsilon\,\delta\ b/r(1+z)= \epsilon\,b/\gamma\, c\, t$
in the observer frame. Hence, the  photon flux 
above the detection threshold $E_{min}$ as seen by the distant observer is 
given by
\begin{equation}
N(t, E>E_{min}) \propto R_{CB}^2 \int_{E_{min}}^\infty \int_0^R
                {dn\over d\epsilon}{d\epsilon\over dE}  
                 2\pi\, b\, db\, dE\,
\label{Eq6}
\end{equation}
which yields the late-time pulse shape in the limit $t\gg\tau$.

\section{The X-ray Afterglow}
The afterglow observations with the Swift X-ray 
telescope (XRT) in the 0.3-10 keV range begins usually 
during the fast decline phase of the prompt emission. 
The energy flux above $E_{min}$ due to ICS during this 
phase is  given similarly by 
\begin{equation}
F(E>E_{min})=\int E{dN\over dE}dE \propto t^{-2}\, exp(-t/\tau(E_{min}))\, . 
\label{Eq7}
\end{equation} 
The E-dependence of $\tau(E)=(R/\gamma\,c)(kT\,E)$ produces the 
fast spectral softening observed during the fast decline phase of the 
promt emission pulses (Evans et al. 2007; 2009). In SN-GRBs, this fast 
decline phase is overtaken by synchrotron radiation from the 
decelerating jet in the relatively high density interstellar 
environment (e.g., a molecular cloud, where most SNeIc usually take place). 
In SN-Less GRBs, the fast decline phase is overtaken by an afterglow 
produced by a plerion powered by a millisecond pulsar which underwent 
a phase transition (to a quark star ?) due to mass accretion in a 
compact binary (Dado \& Dar 2017).

In  a constant density interstellar medium (ISM) this deceleration yields 
(Dado, Dar \& De R\'ujula 2009a; Dado \& Dar 2014)
\begin{equation}
\gamma(t)={\gamma(0)\over [\sqrt{(1+\gamma(0)^2\theta^2)^2+t/t_s}-
          \gamma(0)^2\theta^2]^{1/2}}    
\label{Eq8}
\end{equation} 
and
\begin{equation} 
F(E,t)\propto [\gamma(t)]^{3\beta-1}[\delta(t)]^{\beta+3}\,\nu^{-\beta}
\label{Eq9}
\end{equation} 
where $\beta$ is the spectral index of the X-rays. Eq.(9) yields 
a late-time power-law decline (e.g. Dado \& Dar 2016) 
$F(E,t)\rightarrow \nu^{-\beta}t^{-\beta-1/2}$.

A plerion powered by a pulsar, with an initial period $P(0)$ and period derivative 
$\dot{P}(0)$, produces an afterglow with a light curve $F\propto 
F_{ps}(1+t/t_b)^{-2}$ (Dado \& Dar 2017) where a transition from a plateau 
to $\sim t^{-2}$ decrease takes place around $t_b=P(0)/2\dot{P}(0)$.

In Figures 3 and 4 the light curve of the X-ray afterglow of 
GRB171205A, that was measured with the Swift XRT and reported in the 
Swift-XRT GRB lightcurve repository (Evans et al. 2007,2009)
is compared to the 
CB model best fit lightcurves for a fast declining ICS emission 
(Eq.(7)) taken over the synchrotron emission from a decelerating jet 
in a constant density ISM (Eq.(9)), or by X-ray emission from a 
plerion powered by a pulsar, respectively. The prompt emission involved 
only two adjustable parameters, a normalization and $\tau(Emin)$. The 
synchrotron afterglow involved three  adjustable parameters, a 
normalization, a deceleration time $t_s=426$ s and $\gamma_0\theta= 3.24$. 
The spectral index of the X-ray afterglow, $\beta\approx 0.92$, was that 
measured with Swift. The best fit value $\gamma_0\theta= 3.24$ and 
the viewing angle $\theta=5$ mrad  yielded $\gamma_0=648$ consistent with 
that estimated for ordinary GRBs (Dado \& Dar 2014).

The pulsar powered afterglow fit involved only  
two adjustable parameters, a normalization and $t_b=4.5$ days, which 
yielded $P(0)\sim 145$ ms and $\dot{P}(0)\sim 1.86\times 10^{-7}$  
(Dado \& Dar 2017). 
The two best fits have comparable $\chi^2/dof$, 1.31 and 1.38, 
respectively, which cannot  distinguish yet between 
a jet afterglow and  a pulsar powered afterglow. 
\begin{figure}[]
\centering 
\epsfig{file=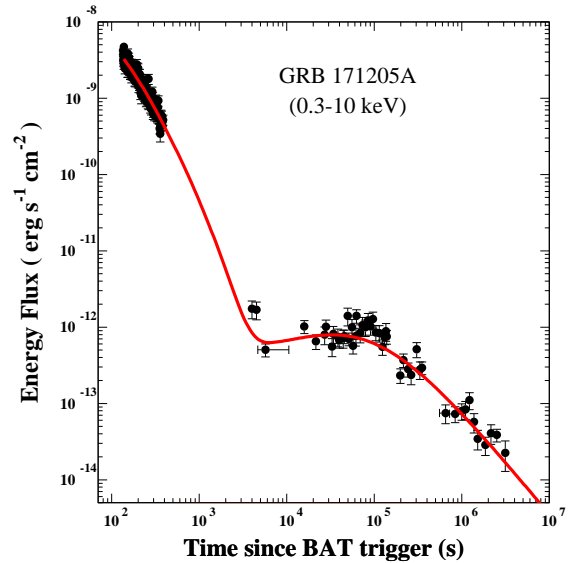,width=8.cm,height=8.cm}
\caption{The light curve of the X-ray afterglow of GRB171205A 
reported in  the
Swift-XRT GRB lightcurve repository (Evans et al. 2007,2009)
compared to the  afterglow predicted by the CB model,  
assuming a far off-axis ordinary SN-GRB.}  
\label{Fig3}
\end{figure}
\begin{figure}[]
\centering 
\epsfig{file=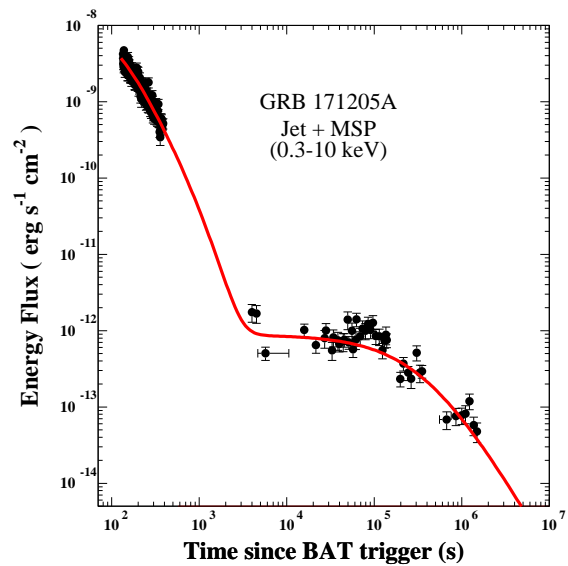,width=8.cm,height=8.cm}
\caption{The light curve of the X-ray afterglow of GRB171205A 
reported in  the
Swift-XRT GRB lightcurve repository (Evans et al. 2007,2009)
compared to the predicted decay by the CB model of its prommpt emission    
taken over by plerion emission powered by a nascent pulsar.}  
\label{Fig4}
\end{figure}
\section{Superluminal Motion}
A very specific prediction of the CB model 
(Dar \& De R\'ujula 2000b; Dado, Dar \& De R\'ujula 2003, 2016)
is an apparent superluminal velocity of the 
afterglow of SN-GRBs in the plane of the sky, which 
is given by 
\begin{equation}
V_{app}\approx {2\,\gamma^2\,c\, \theta \over (1+\gamma^2\theta^2)(1+z)}\,.  
\label{Eq10}
\end{equation} 
In far off-axis SN-GRBs, $V_{app}\approx 2\,c/(1+z)\theta$ as long as 
$\gamma^2\theta^2>>1$, i.e.,  $t<<(1+\gamma_0^2\theta^2)^2\,t_s$,
and $V_{app}\propto t^{-1/2}$ for $t>>(1+\gamma_0^2\theta^2)^2\,t_s$.

In the case of a plerion afterglow, no 
separation between the GRB and its afterglow is expected.
In Figures 5,6 we present a plot of the apparent superluminal velocity
of the afterglow of GRB171205A in the plane of the sky 
and the angular displacement of the afterglow from the burst position, 
respectively, as predicted by the CB model. The
viewing angle $\theta\approx 5 $ mrad 
and the deceleration time scale  $t_s\approx 426$ s are those used to 
reproduce the light curve of  X-ray synchrotron  afterglow shown in 
Figure 3.
\begin{figure}[]
\centering
\epsfig{file= 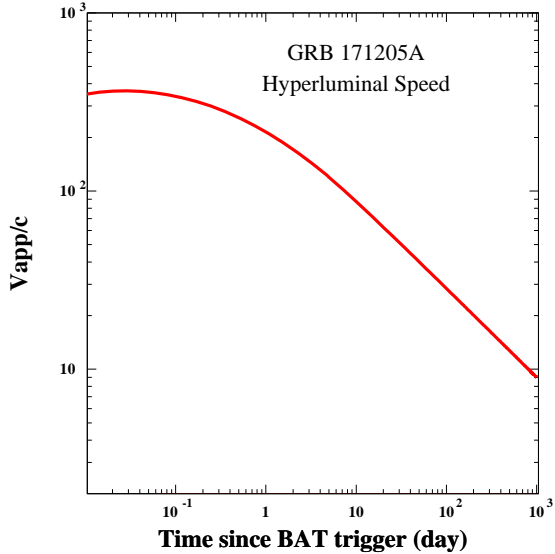,width=8.cm,height=8.cm}
\caption{The expected superluminal velocity of the highly 
relativistic jet, which produced the LL GRB171205A, 
as function of time, based on the parameters of the best fit 
CB model lightcurve to its X-ray afterglow.}
\label{Fig5}
\end{figure}
\begin{figure}[]
\centering
\epsfig{file= 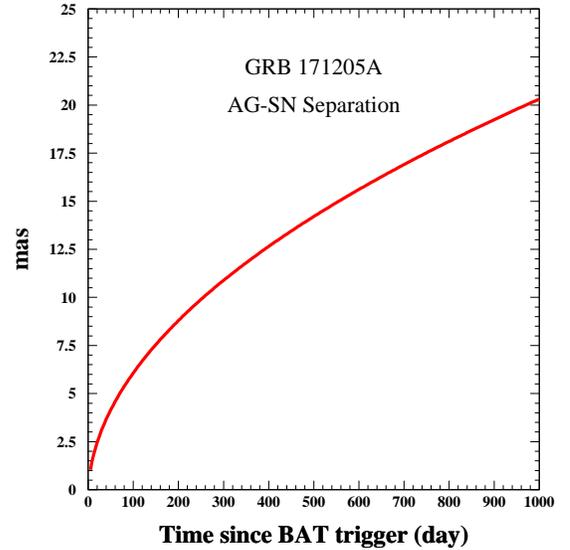,width=8.cm,height=8.cm}
\caption{The angular separation
between the afterglow of GRB171205A and the GRB-SN location as
function of time after burst expected in the CB model from 
the parameters of the highly relativistic jet extracted from  
the lightcurve of the X-ray afterglow of GRB171205.}
\label{Fig6}
\end{figure}

\section{Conclusions}
The observations of the low luminosity GRB171205A and its afterglow, 
analyzed in the framework of the cannonball model of GRBs, indicate that 
it is an ordinary SN-GRB viewed from far off-axis like other nearby low 
luminosity SN-GRBs, which were analyzed by the cannonball model. The 
afterglow of such ordinary SN-GRBs that could be viewed from far off-axis 
because of their proximity, should display a large observable 
hyperrluminal velocity at early time, which decays at late time like 
$t^{-1/2}$ as long as the jet moves within a constant density ISM of its 
host galaxy. Their proximity offers a unique opportunity for high 
resolution VLA, VLBA, and VLBI follow-up radio observations of those with 
a bright radio afterglow, such as GRB171205A, to measure their 
apparent superluminal motion. This will test whether GRB171205A  and 
probably most other nearby low luminosity SN-GRBs, are ordinary SN-GRBs 
viewed from far off-axis, as advocated by the cannonball model of GRBs,
or belong to a different class of SN-GRBs, which 
cannot be accommodated together with ordinary SN-GRBs in the standard 
fireball models of GRBs.

\end{document}